\documentclass[twocolumn,preprintnumbers,amsmath,amssymb,floatfix]{revtex4}
\usepackage{graphicx}
\usepackage{dcolumn}
\usepackage{bm}
\graphicspath{{./},{./figs/}}
\usepackage[latin1]{inputenc}

\begin{document}


\title{Excitonic gaps and exciton binding energies in multilayer phosphorene quantum dots}

\author{M. A. Lino$^{1,2}$, J. S. de Sousa$^1$,  D. R. da Costa$^1$, A. Chaves$^1$, J. M. Pereira Jr.$^1$, G. A. Farias$^1$}
\affiliation{$^1$Universidade Federal do Cear\'a, Departamento de F\'{\i}sica, Caixa Postal 6030, 60455-760 Fortaleza,  Cear\'a, Brazil}
\affiliation{$^2$Universidade Federal do Piau\'{\i}, Departamento de F\'{\i}sica, CEP 64049-550, Teresina, Piau\'{\i}, Brazil}


\begin{abstract}
Dielectric screening is greatly important to an accurate calculation of the exciton binding energies in two-dimensional materials. In this work, we calculate the dielectric function and 2D polarizability of multilayer (up to three) phosphorene sheets using Density Functional Theory. The 2D polarizabilities are then used in the dielectric screening of the excitonic interaction in multilayer phosphorene quantum dots. In the limit of large quantum dots, excitonic gaps are shown to exhibit very good agreement with state-of-the-art measurements of the optical gaps of multilayer phosphorene sheets deposited in different substrates. 
\end{abstract}

\maketitle



Few-layer black phosphorus (BP), also known as phosphorene, is a two-dimensional (2D) material exhibitng interesting physical and chemical properties. Just like graphene, it can be exfoliated to an atomically thin sheet of phosphorus. But differently from graphene, it exhibits a direct band gap inversely proportional to the number of layers $N$, varying from 0.35 eV ($N\rightarrow \infty$) to a band gap of the order of 2 eV ($N = 1$). More specifically, the single-particle gap of bare monolayer BP has been reported as varying between $1.52$ eV and $2.12$ eV \cite{xia2014,castellanos2015,liang2014, rudenko2015, zhang2016, li2017}, but most of these studies did not take into account neither excitonic nor substrate effects. However, it was recently shown that BP exhibits very large exciton binding energies that are able to withstand large in-plane electric fields, and are prone to allow for observation of excited excitonic states \cite{chaves2015,chaves2016}. Furthermore, excitonic effects are particularly interesting in BP because of the highly anysotropic band structure of this material.


Excitonic effects in 2D materials are currently a matter of debate in the literature since the effective Coulomb interaction in purely planar structures is different from the interaction in three dimensions \cite{cudazzo2011,berkelbach2013,rodin2014,latini2015}. For bulk semiconductors, the macroscopic dielectric constant is defined as the limiting value of $\varepsilon(q)$ for $q\rightarrow 0$. For 2D systems, the dielectric screening is non-local assuming the general form $\varepsilon_{2D}(\vec{q}) = 1 + 2\pi \alpha_{2D}(q)$, where $\alpha_{2D}$ is the polarizability of the 2D material. The real space Coulomb interaction assuming such a non-local kind of screening renders a Keldysh-like interaction potential, which was originally derived to describe the electrostatic interaction in thin films with finite thicknesses \cite{keldysh1979}. The thickness of atomically thin 2D materials is not well defined and new models need to be developed for such systems. In this sense,  Cudazzo \textit{et al.} demonstrated that this kind of potential introduces a length scale $r_0 = 2 \pi \alpha_{2D}$ (for an isolated 2D sheet in vacuum) that depends on the polarizability of the planar material \cite{cudazzo2011}. Rodin \textit{et al.} improved Cudazzo's model by including the effect of the substrate and showing that  $r_0 = 2 \pi \alpha_{2D}/\kappa$ (for a 2D sheet sandwiched between vacuum and a substrate with dielectric constant of $\varepsilon_{sub}$), where $\kappa = (1+\varepsilon_{sub})/2$ \cite{rodin2014}. Taking advantage of this simple formulation, Cudazzo \textit{et al.}, Berkelbach \textit{et al.} and Rodin \textit{et al.} studied excitonic interactions in monolayer graphane, transition metal dichalcogenides and phosphorene, respectively \cite{cudazzo2011,berkelbach2013,rodin2014}. They all used \emph{ab initio} methods to determine the 2D polarizability of each material. Later on, Latini \textit{et al.}  extended those studies by including the nearly vanishing thickness of the 2D materials to study excitons in  hexagonal boron-nitride MoS$_2$ monolayers \cite{latini2015}. Their results  shown that the strict 2D model provides very good results of exciton binding energies as compared to their quasi-2D formulation.

Recently, the authors used the dielectric screening model of Rodin \textit{et al.} to investigate the role of different substrates in the excitonic fine structure of monolayer BP quantum dots (BPQDs) \cite{desousa2017}. They have shown that the excitonic gaps are well described by a sum of power laws $E_X(R) = E_g^{(bulk)} + A/ R^{n} - B/R^{m}$ where $R$ is the QD radius, and $A$, $B$, $C$, $\gamma$, $n$, and $m$ are substrate-dependent parameters. Our previous results converged to the experimental optical gaps of monolayer BP deposited on different substrates for $R \rightarrow \infty$  \cite{zhang2016, li2017}. As the screening model was developed for a single BP layer, we could not investigate the excitonic properties of multilayer BPQDs. 

In this work, we extend our previous study by calculating the size-dependent excitonic gaps and exciton binding energies of (up to three) multilayer BPQDs,  using a dielectric screening model parameterized by \emph{ab initio} calculations. Although multilayer sheets are thicker than the monolayer, we modeled them as a strictly 2D material, and calculated their 2D polarizabilities as a function of the number of layers following the recipe of Rodin \textit{et al.} \cite{rodin2014}. In the limit of large QDs, our results reproduced very well the experimental optical gaps of multilayer BP deposited on different substrates.


Multilayer circular BPQDs were formed by generating a large BP sheet with a given number of layers (up to three assuming AB stacking) sheet with armchair (zigzag) direction aligned to the $x$ ($y$) axis, and the atoms outside a given radius $R$ (measured with respect to center of mass of the large sheet) were disregarded. The resulting circular BPQDs exhibit a mixture of different types of edges at their boundaries. The energy spectrum of the BPQDs was calculated by solving Schroedinger equation represented in a linear combination of atomic orbital (LCAO) basis, such that the effective Hamiltonian reads $\hat{H}  = \sum_i \epsilon_i |i\rangle\langle i | + \sum_{i,j} t_{i,j} |i\rangle\langle j |$. The generalized index $i = \{\vec{R}_i,\alpha,\nu\}$ represents the orbital $\nu$ of the atomic species $\alpha$ at the atomic site $\vec{R}_i$, $\epsilon_i$ represents the onsite energy of the i-th site, and $t_{i,j}$ represents the hopping parameter between i-th and j-th sites. As for the hopping parameters and lattice constants, we adopted the $10$ hopping parameter TB model of Rudenko \textit{et al.} \cite{rudenko2015}.

The size-dependent excitonic gaps $E_X(R)$ are calculated perturbatively as $E_X(R) = E_g(R) - E_B(R)$, where $E_g(R) = \varepsilon_{cbm}(R) - \varepsilon_{vbm}(R)$ is the single-particle gap, $\varepsilon_{cbm}$  and $\varepsilon_{vbm}$ represent the conduction band minimum and valence band maximum energies, respectively. The exciton binding energy  $E_B(R)$ is calculated as

\begin{eqnarray}
\label{eq:eb} 
E_B=\int d\vec{r}_1 d\vec{r}_2  |\psi_{cbm}(\vec{r}_1)|^2  V(|\vec{r}_1-\vec{r}_2 |)  |\psi_{vbm}(\vec{r}_2)|^2 ,
\end{eqnarray}

\noindent where $\psi_{cbm}$ and $\psi_{vbm}$  represent the wavefunctions of states corresponding to the conduction band minimum and valence band maximum, respectively.

\noindent The electron-hole interaction potential  $V(|\vec{r}_1-\vec{r}_2 |)$ is given by:
\begin{equation}
\label{eq:dielectric}
V(r) = \frac{q^2}{4\pi\varepsilon_0} \frac{\pi}{2\kappa r_0} \left[ H_0\left( \frac{r}{r_0}\right)-Y_0\left( \frac{r}{r_0}\right)\right],
\end{equation}
\noindent where we adopted the model of Rodin \textit{et al.} for the Coulomb interaction between charges confined in a two-dimensional material sandwiched between a substrate with dielectric constant $\varepsilon_{sub}$ and vacuum \cite{rodin2014}. In the above expression, $r$ is the distance between particles,  $r_0 = 2\pi \alpha_{2D}/\kappa$, $\kappa = (1+ \varepsilon_{sub})/2$,  $H_0$ and $Y_0$ are the Struve and Neumann functions, and $\alpha_{2D}$ represents the 2D polarizability of the multilayers. This quantity is obtained following the method described by Berkelbach \textit{et al.} \cite{berkelbach2013}, who calculated the real component of static dielectric permittivity $\varepsilon$ as a function of the interlayer distance $d$ of a single BP sheet:

\begin{equation}
\label{eq:eps}
\varepsilon = 1 + \frac{4\pi \alpha_{2D}}{L_z}.
\end{equation}

\noindent In our multilayer calculations, we adopted $L_z$ as the unit cell size in z direction (perpendicular to multilayer sheets) and assumed it as being large enough to prevent interaction among BP sheets and their multiple copies imposed by periodic boundary conditions. 

The dielectric function of multilayer BP sheets was calculated using the Density Functional Theory (DFT) as implemented in SIESTA code \cite{soler96,soler2002}. We make use of the Generalized Gradient Approximation (GGA) for the exchange-correlation functional \cite{perdew96} and norm-conserving Troullier-Martins pseudopotentials \cite{troullier91} in Kleinman-Bylander factorized form \cite{Kleinman82}. We used double-zeta basis set (DZP) composed of numerical atomic orbitals of finite range augmented by polarization functions. The fineness of the real-space grid integration was defined by a minimal energy cutoff of 180 Ry. The range of each orbital is determined by an orbital energy confinement of 0.01 Ry. The geometries were considered optimized when the residual force components were less than 0.04 eV/\AA. Due to the well known problem of gap understimation of DFT, we applied the scissors operator such that the single particle gap as function of the number of layers reflected the values obtained by the GW calculations of Rudenko \textit{et al.} \cite{rudenko2015}.

\begin{figure}[ht]
\begin{center}
\includegraphics[width=.5\textwidth,clip=true]{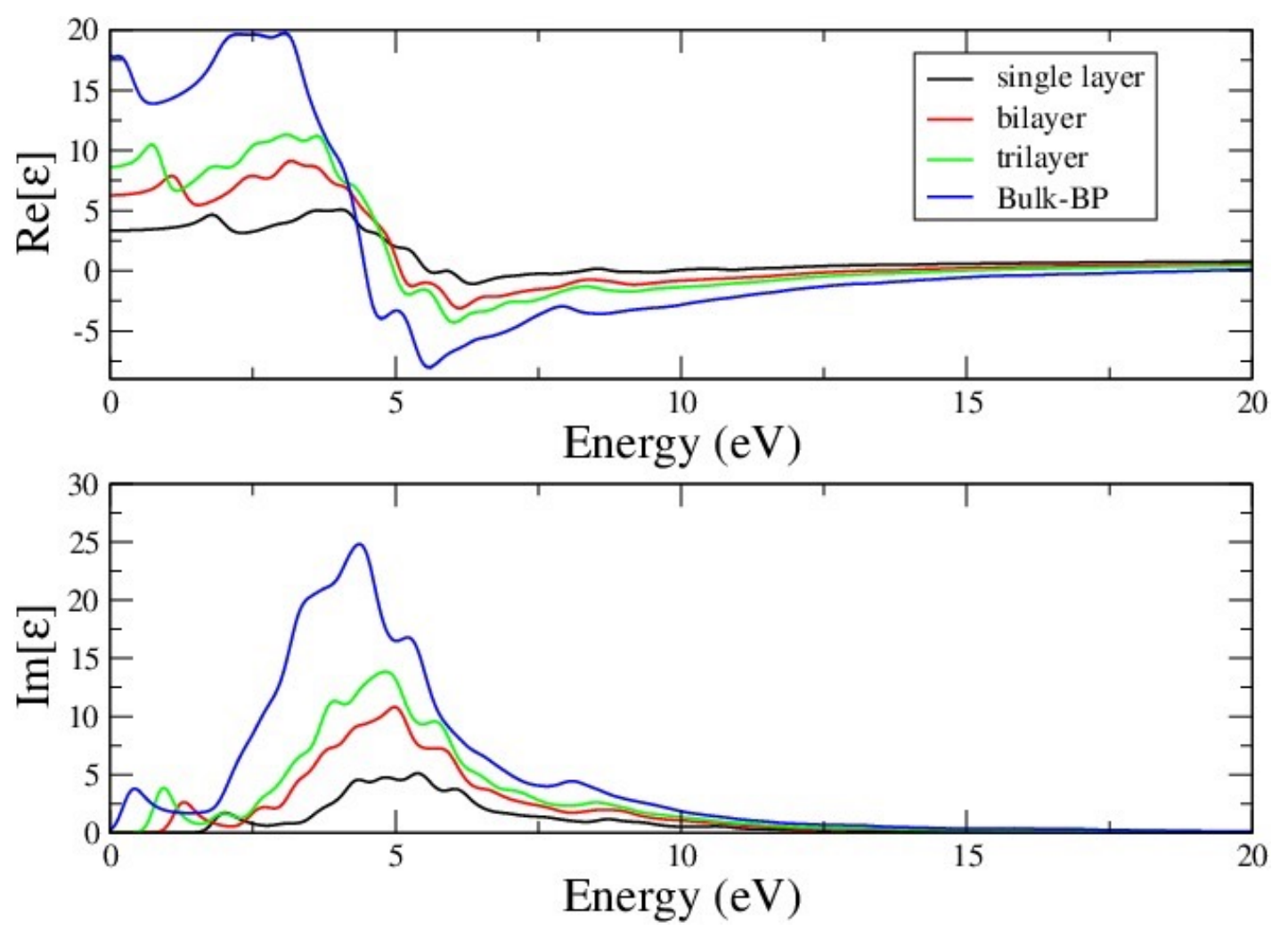}
\caption{\label{fig:eps} Real and imaginary components of the dielectric function of multilayer BP.}
\end{center}
\end{figure}

Figure \ref{fig:eps} shows the real and imaginary components dielectric function of BP sheets up to three layers.  The real component of the dielectric constant of BP increases with the number layers. The 2D polarizability as function of the number of layers is obtained applying Eq.~\ref{eq:eps} for $Re[\varepsilon(0)]$. These results are shown in Table \ref{tab:param}. Our monolayer calculation resulted in $\alpha_{2D} = 4.72$ \AA, which is in good agreement with $\alpha_{2D} = 4.1$ \AA~calculated by Rodin \textit{et al.} \cite{rodin2014}. Kumar et al.  also investigated the polarizability of multilayer BP by means of DFT calculations. However, their results cannot be directly compared to ours since they used a different definition of slab polarizability \cite{kumar2016}.

\begin{figure}[ht]
\begin{center}
\includegraphics[width=.45\textwidth,clip=true]{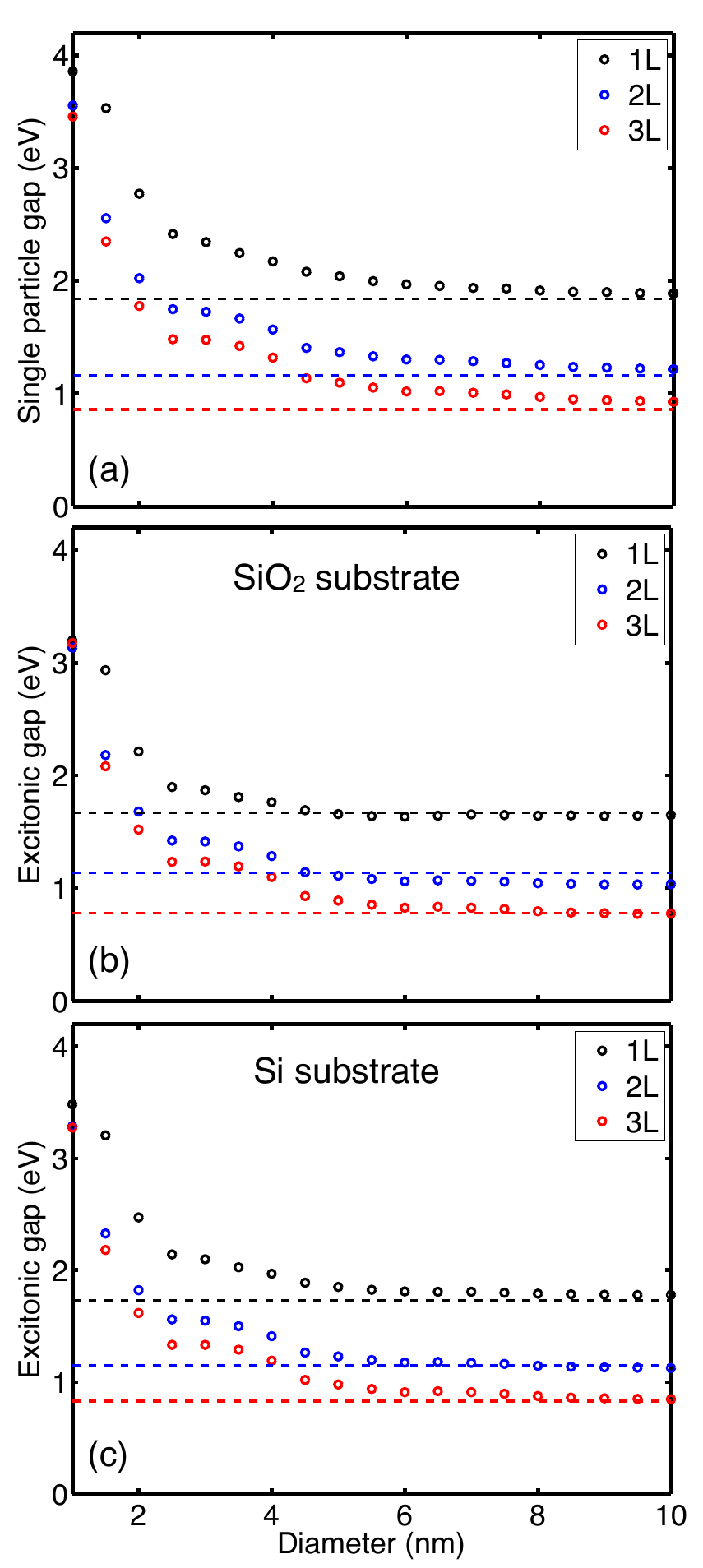}
\caption{\label{fig:exc} Size-dependent (a) single particle gaps and excitonic gaps of multilayer BPQDs deposited on (b) SiO$_2$ and (c) Si substrates. The horizontal lines in (a) represent the single-particle gaps of bulk multilayer BP. In (b) and (c), the horizontal lines represent the excitonic gaps of bulk multilayer BP measured by Zhang et al. \cite{zhang2016} and Li et al. \cite{li2017}, respectively.}
\end{center}
\end{figure}

Figure \ref{fig:exc} shows the size-dependent single particle gaps and excitonic gaps of multilayer BPQDs deposited on SiO$_2$ ($\varepsilon_{sub} = 3.9$) and Si ($\varepsilon_{sub} = 11.7$).  The size-dependent TB calculations converge to the single-particle gaps of the infinite sheets in the limit of large BPQDs, and the 10 nm wide BPQD is demonstrated to be large enough to exhibit nearly bulk properties. Since the size-dependent experimental optical gaps of BPQDs have not been reported yet, we will compare our calculations with a couple of  experimental studies on large BP sheets with multiple number of layers. In this regard, Zhang and Li measured the optical gaps of BP as a function of the number of layers deposited on substrates with dielectric constants similar to the ones of SiO$_2$ and Si, respectively \cite{zhang2016,li2017}. TB calculations have shown that the size-dependent excitonic gaps of BPQDs in SiO$_2$ are slightly lower than the ones of BPQDs on Si. This was also observed when comparing the measurements of Zhang and Li for any number of BP layers, but this difference decreases as the number of BP layers increases.  

Figure \ref{fig:exc}(b)-(c) shows that, in the limit of large BPQDs ($R = 5$ nm), the TB calculations of the excitonic gaps converge to the measured bulk values of Zhang and Li for all number of BP layers and substrates with dielectric constants as distinct as SiO$_2$ and Si. The calculated electron-hole interaction based on the 2D model of Rodin \textit{ et al.} \cite{rodin2014} and parameterized by \emph{ab initio} calculations of the 2D polarizabilities $\alpha_{2D}$ as a function on the number of BP layers reveals that the major contribution of the difference of optical gaps measured by Zhang and Li is due to electrostatic coupling among the in-plane electron-hole pair and the substrate \cite{zhang2016,li2017}. 

\begin{figure}[ht]
\begin{center}
\includegraphics[width=.45\textwidth,clip=true]{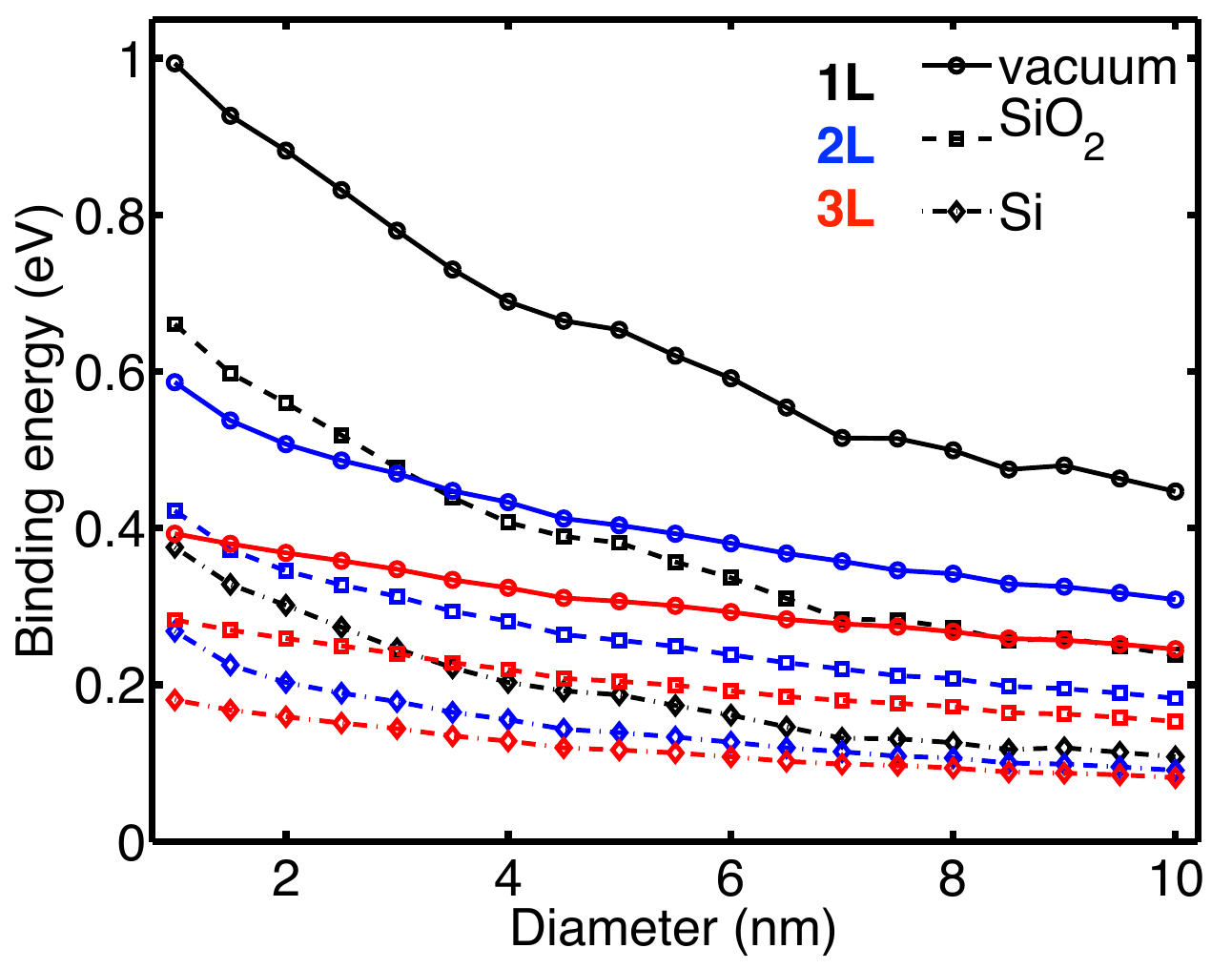}
\caption{\label{fig:eb} Size-dependent exciton binding energies of multilayer BPQDs in vacuum (solid lines), as well as over SiO$_2$ (dashed lines) and Si (dot-dashed lines) substrates.}
\end{center}
\end{figure}

The size-dependent electron-hole binding energies of BPQDs with different number of layers is shown in Figure \ref{fig:eb}. Here, it is clear the dramatic effect of the substrate and the number of layers in the excitonic gaps. For large monolayer QDs ($R = 5$ nm) the binding energy decreases from approximately 0.45 eV (in vacuum) to 0.24 eV (on SiO$_2$) and 0.11 eV (on Si). For bilayer QDs, the binding energy decreases from approximately 0.30 eV (in vacuum) to 0.18 eV (in SiO$_2$) and 0.07 eV (on Si). The comparison of the excitonic gaps and binding energies calculated in this work for large multilayer QDs and the measured optical gaps of Zhang and Li is shown in Table \ref{tab:param}.

In conclusion, we have studied the excitonic gaps of multilayer BPQDs (up to $N = 3$) using a dielectric screening model parametrized by \emph{ab initio} calculations. In general, we obtained very good agreement between theoretical and experimental data, with the largest discrepancy obtained for the bi-layer QDs in SiO$_2$, where our estimated excitonic gap was 0.1 eV below the experimental gap of the infinite BP bilayer. Subtle effects like the single-particle gap broadening due to coupling of carriers in BP and substrate polarons  were neglected because of their small amplitude. As described by Mogulkoc \textit{et al.} \cite{mogulkoc2016}, the single-particle gap of monolayer BP deposited on SiO$_2$ is enlarged by 30 meV. If those effects we considered, the agreement between experiments and our calculations would be further improved. Finally, we emphasise that the strong dependence of the excitonic gaps  on the substrate and number BP layers opens up the possibility of tailoring the optical properties of BPQDs by size and substrate engineering.

\begin{table*}[htp]
\caption{\label{tab:param} BP parameters calculated for different numbers of BP layers. }
\begin{center}
\begin{tabular}{crrrrcccc}
\hline\hline 
$N$ & $L_z$(\AA) & $Re[\varepsilon(0)]$ & $\alpha_{2D}$(\AA)& $E_G$(eV) & $E_X^{SiO_2}$(eV) & $E_b^{SiO_2}$(eV) & $E_X^{Si}$(eV) & $E_b^{Si}$(eV) \\ \hline \hline
1     & 25.26 & 3.35  & 4.72    & 1.84$^{a}$ & 1.67$^{c}$ & 0.17$^{e}$ & 1.73$^{d}$ &  0.11$^{e}$ \\
\multicolumn{3}{c}{}  & 4.10$^{f}$    & 1.89$^{b}$ & 1.65$^{b}$ & 0.24$^{b}$  & 1.78$^{b}$  &   0.11$^{b}$ \\ \hline
2     & 25.32 & 6.28  & 10.64  & 1.16$^{a}$  &1.14$^{c}$ & 0.02$^{e}$ & 1.15$^{d}$ & 0.01$^{e}$\\
\multicolumn{4}{c}{}                & 1.22$^{b}$ & 1.04$^{b}$ & 0.18$^{b}$  & 1.15$^{b}$  &   0.07$^{b}$ \\ \hline
3     & 28.92 & 8.64  & 17.59  & 0.86$^{a}$ & 0.78$^{c}$ & 0.08$^{e}$ & 0.83$^{d}$ & 0.03$^{e}$ \\
\multicolumn{4}{c}{}                & 0.95$^{b}$ & 0.78$^{b}$  & 0.17$^{b}$  & 0.85$^{b}$  &   0.10$^{b}$ \\ \hline\hline 
\end{tabular}
\end{center}
$^{a}$TB calculation for infinite layers \\ $^{b}$TB calculation for the largest QDs ($R = 5$ nm)\\ $^{c}$exp. \cite{zhang2016} \\ $^{d}$exp. \cite{li2017}\\  $^{e}$estimated as $E_b = E_G^{(bulk)} - E_X^{(exp.)} $ \\ $^{f}$DFT calculation \cite{rodin2014}
\end{table*}%




\end{document}